\documentclass[10pt]{article}
\usepackage{geometry}                
\usepackage{amsmath}
\usepackage{graphicx}
\usepackage{dcolumn}
\usepackage{bm}
\usepackage{wrapfig}
\usepackage{cite}
\usepackage{epsfig}
\usepackage{breqn}
\usepackage{amssymb}
\usepackage{subfigure}
\usepackage[usenames,dvipsnames]{pstricks}
\usepackage{epsfig}
\usepackage{pst-grad} 
\usepackage{pst-plot} 
 \usepackage[usenames,dvipsnames]{pstricks}
 \usepackage{epsfig}
 \usepackage{pst-grad} 
 \usepackage{pst-plot} 
 \usepackage{float}

\begin{document}

\title{The Osgood Criterion and Finite-Time Cosmological Singularities}
\author{Ikjyot Singh Kohli \\isk@mathstat.yorku.ca \\York University - Department of Mathematics and Statistics}
\date{\today}                                           

\maketitle

\begin{abstract}
In this paper, we apply Osgood's criterion from the theory of ordinary differential equations to detect finite-time singularities in a spatially flat FLRW universe in the context of a perfect fluid, a perfect fluid with bulk viscosity, and a Chaplygin and anti-Chaplygin gas. In particular, we applied Osgood's criterion to demonstrate singularity behaviour for Type 0/big crunch singularities as well as Type II/sudden singularities. We show that in each case the choice of initial conditions is important as a certain number of initial conditions leads to finite-time, Type 0 singularities, while other precise choices of initial conditions which depend on the cosmological matter parameters and the cosmological constant can avoid such a  finite-time singularity. Osgood's criterion provides a powerful and yet simple way of deducing the existence of these singularities, and also interestingly enough, provides clues of how to eliminate singularities from certain cosmological models. 
\end{abstract}


\section{Introduction}
The Osgood criterion \cite{osgood2} is a classical criterion, due to W.F. Osgood in 1898 which gives conditions for ordinary differential equations to admit unique solutions and also to have singularities where solutions explode in a finite time. Following the conventions in \cite{leonja}, Osgood's criterion states  a solution $x(t)$ of the initial value problem
\begin{eqnarray}
\label{eq:ivp1}
\dot{x} &=& f \left[x(t)\right], \\
\label{eq:ivp2}
x\left(t_{0}\right) &=& \xi,
\end{eqnarray}
blows up in finite time if and only if
\begin{equation}
\label{eq:osgoodcond1}
\int_{\xi}^{\infty} \frac{ds}{f(s)} < \infty.
\end{equation}

This condition can be derived from Barrow's formula \cite{arnoldode}, which says that the solution $\phi(t)$ of such an ODE is given by solving
\begin{equation}
t - t_{0} = \int_{\xi}^{\phi(t)} \frac{ds}{f(s)}.
\end{equation}
Suppose the solution becomes singular in a finite time, $t_{s}$, that is, $\phi(t_{s}) = \infty$, where $t_{s} < \infty$, then we have that:
\begin{equation}
t_{s} = \int_{\xi}^{\infty} \frac{ds}{f(s)} + t_{0} < \infty.
\end{equation}

Further, a solution to Eqs. \eqref{eq:ivp1}-\eqref{eq:ivp2} is unique if
\begin{equation}
\int_{\xi}^{\phi(t)} \frac{ds}{f(s)} = \infty.
\end{equation}

Certainly, this is a criterion that is of tremendous importance for physical applications, and as we show in this paper, it is of particular importance in cosmology. In cosmology, assumptions of a spatially homogeneous spacetime, allow Einstein's field equations to be written as an autonomous system of ordinary differential equations. Divergent solutions, that is, solutions that ``blow up'' in a finite time are indicative of finite-time physical and curvature singularities that occur in universe models. As we show below, Osgood's criterion provides a powerful and yet simple way of deducing the existence of these singularities, and also interestingly enough, provides clues of how to eliminate singularities from certain cosmological models. Despite this, we note that Osgood's criterion has received very little attention in the physics and cosmology community. For this reason, we believe that the work here is unique, and could be of interest in future cosmological studies.

In this paper, we will attempt to study the question of whether an expanding spatially flat Friedmann-Lema\^{i}rte-Robertson-Walker (FLRW) universe with cosmological constant and ordinary and exotic matter will expand forever in the future or develop a finite-time singularity. It is understood that the ultimate fate of the universe depends on possible decay of dark energy in the future, and if it does not occur, the universe will expand forever \cite{elliscosmo}. 

Further, the concept of the existence of singularities in the context of general relativity has been studied quite extensively. With respect to the cosmological case, there is the fundamental singularity theorem \cite{elliscosmo} \cite{tolmanward} \cite{raych1955} which describes irrotational geodesic singularities, and states that if $\Lambda \leq 0$, $\mu + 3p \geq 0$, and $\mu + p > 0$ in a fluid flow where $\mu$ is the energy density of the fluid, $p$ is the pressure of the fluid, in addition to having $\dot{u} = 0$, $\omega = 0$, and $H_{0} = 0$ at some time $s_{0}$, then a spacetime singularity, where either the expansion scalar goes to zero or the shear scalar diverges occurs at a finite proper time $\tau_{0} \leq 1/H_{0}$ before $s_{0}$.

Further elaborations are built upon this singularity theorem. It is of interest to note that in \cite{elliscosmo} five possible routes to avoid the conclusions of this singularity theorem are discussed in detail. They are a positive cosmological constant, acceleration, vorticity, an energy condition violation, or alternative gravitational equations. We refer the interested reader to \cite{elliscosmo} for further discussions regarding these issues. 

Following \cite{elliscosmo}, we also note that singularities occur in cosmology not only in the context of FLRW models, but also for realistic anisotropic and inhomogeneous models of the universe in which the strong energy condition $\mu + 3p > 0$ is satisfied. Related to this, Penrose \cite{penrose1965} did pioneering work on black hole singularities producing a theorem that allowed one to predict the existence of singularities in realistic gravitational collapse cases. Hawking \cite{penrosehawking1970} extended these results to the cosmological context leading to the famous Hawking-Penrose singularity theorem which implied that space-time singularities are to be expected if either the universe is spatially closed or there is an `object' undergoing relativistic gravitational collapse (existence of a trapped surface) or there is a point $p$ whose past null cone encounters sufficient matter that the divergence of the null rays through $p$ changes sign somewhere to the past of $p$ (i.e. there is a minimum apparent solid angle, as viewed from $p$ for small objects of given size). The theorem applies if the following four physical assumptions are made: (i) Einstein's equations hold (with zero or negative cosmological constant), (ii) the energy density is nowhere less than minus each principal pressure nor less than minus the sum of the three principal pressures (the `energy condition'), (iii) there are no closed timelike curves, (iv) every timelike or null geodesic enters a region where the curvature is not specially aligned with the geodesic. (This last condition would hold in any sufficiently general physically realistic model.) 

Further, Bekenstein \cite{bekenstein1974} studied exact solutions of Einstein-conformal scalar equations and presented a class of FLRW models which contained both incoherent radiation and a homogeneous conformal scalar field which bounce and never pass through a singular state, thereby circumventing the singularity theorems by violating the energy condition. Parker and Fulling \cite{parkerfulling1973} considered a classical gravitational field minimally coupled to a quantized neutral scalar field possessing mass. They concluded that quantum effects can sometimes lead to the avoidance of the cosmological singularity, at least on the time scale of one Friedmann expansion. Collins and Ellis \cite{collinsellis1979} examined in detail the singularities that occur in Bianchi cosmologies.
Barrow and Matzner \cite{barrowmatz1977} showed that the singularity corresponding to homogeneous and isotropic universes observationally equivalent to ours must be of simultaneous Robertson-Walker type containing only small curvature fluctuations. In an interesting application of Robertson-Walker singularities, Barrow \cite{barrow1977} showed that the large-scale velocity and vorticity fields required for the vortex and spinning-core theories of galaxy formation can be generated primordially in a natural way from a FLRW singularity.

Related to future \emph{finite-time} singularities that develop in FLRW models is the question of the so-called ``ultimate fate of the universe'', there are several possibilities \cite{2014PhRvD..90f4014F} which we list in Table \ref{table:table1}.

\begin{table}[h]
\centering
\begin{tabular}{|c| |c| |c| |c|}
\hline
\emph{Singularity Type} & \emph{Description} & \emph{T} & \emph{K} \\
\hline
Big Crunch (Type 0) & $\theta, p, \mu \to \infty, l \to 0$ & Strong & Strong \\ \hline
Big Rip (Type I) \cite{phantom1} \cite{2006PhRvD..74f4030F} & $l, p, \mu \to \infty$ & Strong & Strong \\ \hline
Sudden Singularities (Type II) \cite{2004CQGra..21.5619B, 2004PhLB..595....1N, 2004CQGra..21L..79B, 2004CQGra..21L.129L, 2004PhRvD..70j3522N, 2005PhRvD..71j3505D, 2005PhLB..625..184D, 2004MPLA...19.2479C, 2008PhRvD..78l3508B, 2009PhRvD..80d3518B, 2008PhRvD..78d6006N, 2010CQGra..27p5017B, 2012PhRvD..85h3527D, 1986MNRAS.223..835B,  2002CQGra..19L.101S, 2004PhRvD..69l3512G, 2010JCAP...05..034B, 2004PhRvD..70l1503F, 2010PhRvD..82l3534K, 2013PhRvD..88b3535K,2013PhRvD..88f7301B} & $l, \theta, \mu < \infty, p \to \infty$ & Weak & Weak \\ \hline
Big Freeze (Type III) \cite{2008PhLB..659....1B}  & $l < \infty, \theta, \mu, p \to \infty$ & Weak & Strong \\ \hline
Generalized Sudden / Big Separation (Type IV) \cite{2005CQGra..22.1563B} & $l, \theta, \mu, p < \infty, \dot{p} \to \infty$ & Weak & Weak \\ \hline
$w$-Singularities (Type V) \cite{2005PhRvD..71h4018S,2009PhRvD..79f3521D, 2010PhRvD..82l4004F} & $l <\infty, \mu, p \to 0, w \to \infty$ & Weak & Weak \\ \hline
\end{tabular}
\caption{A classification of known finite-time singularities. $l$ refers to the scale factor in the FLRW metric, $p$ and $\mu$ refer to the pressure and energy density of the matter content in the FLRW spacetime, $w$ is the barotropic index, and $\theta$ is the expansion scalar as defined above.  Note that, sudden singularities are weak singularities where the spacetime can be extended after the singular event. The singularities are classified as being strong or weak according to the classification due to Tipler (T) \cite{1977PhLA...64....8T} and Krolak (K) \cite{1986CQGra...3..267K}.}
\label{table:table1}
\end{table}

Further, so-called infinite-time singularities may also occur as well. These include Type $\infty$ or directional singularities, which are of strong type according to both  \cite{1977PhLA...64....8T} and \cite{1986CQGra...3..267K}. There are also other types of non-singular future behaviour which are also of the infinite-time variety. These include the little rip \cite{2011PhRvD..84f3003F, 2012PhLB..708..204F}, pseudo-rip \cite{2012PhRvD..85h3001F}, and the little sibling of the big rip \cite{2015IJMPD..2450078B}. These are discussed in more detail in \cite{2014msu..book..101D} and \cite{2014PhRvD..90f4014F}.



It is also important to note that conditions describing the avoidance of finite-time singularities have been given in \cite{2004PhLB..595....1N}, \cite{2004PhRvD..70d3539E,  2008JCAP...10..045B, 2012Ap&SS.342..155B, 2005PhRvD..72b3003N}.

In this work, we will consider a spatially homogenous and isotropic $k=0$ FLRW universe with a cosmological constant and different types of matter sources. We will use Osgood's criterion to establish when such models admit finite-time, Type 0 singularities as discussed below. 

\section{The Dynamical Equations}
It is well-known that covariant derivative of the four-velocity vector $u_{a}$ can be decomposed as \cite{elliscargese} \cite{hervik}
\begin{equation}
u_{a;b} = \frac{1}{3}\theta h_{ab} + \sigma_{ab} + \omega_{ab} - \dot{u}_{a} u_{b},
\end{equation}
where $u_a$ is the fluid four-velocity, $\theta$ is the expansion scalar, $\sigma_{ab}$ is the shear tensor, $\omega_{ab}$ is the vorticity tensor, and $h_{ab}$ is the standard projection tensor.
For spatially homogeneous universes, we can choose the four-velocity vector to be orthogonal to the space-like surfaces in the $3+1$-spacetime decomposition. With this, the shear tensor $\sigma_{ab}$, the vorticity tensor $\omega_{ab}$, and acceleration $\dot{u}_{a}$ all vanish. One then takes projected, symmetric, and trace-free components of the decomposition equation above in combination with the Einstein field equations
\begin{equation}
R_{ab} - \frac{1}{2}Rg_{ab} + \Lambda g_{ab} = \kappa T_{ab}, 
\end{equation}
to obtain the Raychaudhuri equation
\begin{equation}
\label{eq:raych1}
\dot{\theta} + \frac{1}{3}\theta^2 + \frac{\kappa}{2} \left(\mu + 3 p\right) - \Lambda = 0,
\end{equation}
the energy density evolution equation
\begin{equation}
\label{eq:endensity1}
\dot{\mu} + \theta \left(\mu + p\right) = 0,
\end{equation}
and the generalized Friedmann constraint equation
\begin{equation}
\label{eq:friedmann1}
\frac{1}{3} \theta^2 = \kappa \mu + \Lambda,
\end{equation}
where $\mu$ denotes the matter energy density and $p$ denotes the matter pressure. 

In fact, we have that by the Friedmann equation \eqref{eq:friedmann1}, 
\begin{equation}
\lim_{t \to t_{s}} \theta_{t} = \infty \Rightarrow \lim_{t \to t_{s}} \frac{1}{3} \theta^2 = \infty \Rightarrow \lim_{t \to t_{s}} \left[\kappa \mu(t) + \Lambda\right] = \infty \Rightarrow \lim_{t \to t_{s}} \mu(t) = \infty.
\end{equation}
Therefore, the Osgood criterion provides a way of testing for a singularity where $\theta$ and $\mu$ go to infinity in a finite time. This type of singularity is a big crunch or Type 0 singularity. In the context of singularity classification, the Type 0 singularity is considered a strong singularity \cite{1977PhLA...64....8T, 1986CQGra...3..267K, 2014msu..book..101D}.


The advantage of considering spatially flat models is that the dynamical equations \eqref{eq:raych1}, \eqref{eq:endensity1} can be reduced to a single dynamical equation via Eq. \eqref{eq:friedmann1}. We will therefore consider the dynamics of such models to be represented by an evolution equation for $\theta$, although, the analysis in what follows is equally as valid for an evolution equation for $\mu$. Substituting Eq. \eqref{eq:friedmann1} into Eq. \eqref{eq:raych1}, we obtain
\begin{equation}
\label{eq:raych2}
\dot{\theta} = -\frac{1}{2} \left(1+w\right) \left(\theta^2 - 3 \Lambda\right), 
\end{equation}
where $w$ is the equation of state parameter.

\section{The Existence of Finite-Time Singularities}
With the main evolution equation \eqref{eq:raych2} in hand, we now demonstrate the existence of finite-time singularities. To accomplish this, we will make use of Osgood's criterion as defined earlier.
%

To apply Osgood's criterion to our problem, let us consider the initial value problem
\begin{eqnarray}
\label{eq:raychnew1}
\dot{\theta} &=& -\frac{1}{2} \left(1+w\right) \left(\theta^2 - 3 \Lambda\right), \\
\label{eq:raychnew11}
\theta\left(t_{0}\right) &\equiv& \theta_{0}.
\end{eqnarray}


Following the introductory section and evaluating the integral in \eqref{eq:osgoodcond1}, we obtain
\begin{eqnarray}
t_s - t_{0} &=& \int_{\theta_{0}}^{\infty} -\frac{2 ds}{\left(1+w\right) \left(s^2 - 3 \Lambda\right)} \nonumber \\
\label{eq:integ1}
&=& \frac{1}{\sqrt{3}(1+w)\sqrt{\Lambda}} \log \left| \frac{\theta_0 - \sqrt{3}\sqrt{\Lambda}}{\theta_0 + \sqrt{3} \sqrt{\Lambda}}      \right|.
\end{eqnarray}
Note that $t_s \to \infty$ only if $\theta_{0} = \pm \sqrt{3} \sqrt{\Lambda}$, where the positive value is precisely the value of the de Sitter expansion. Therefore, carefully choosing the initial condition to be equal to the rate of the de Sitter expansion avoids a finite-time singularity where $\theta \to \infty$. 

The value of Eq. \eqref{eq:integ1} is finite and \emph{real} if either 
\begin{equation}
\{\Lambda > 0\} \cap \{\theta_{0} > \sqrt{3} \sqrt{\Lambda}\}.
\end{equation}
Hence, for choices of $\theta_{0}$ that satisfy these inequalities, the  integral in Eq. \eqref{eq:integ1} is finite, and hence $\theta \to \infty$ in a finite time. 

For the sake of completeness, let us consider the case where $\Lambda < 0$. In this case, the integral Eq. \eqref{eq:integ1} is evaluated to be
\begin{equation}
t_{s} - t_{0} = \frac{\pi  \left(-\frac{1}{\Lambda }\right)^{3/2} \Lambda ^{3/2}-2 \tanh ^{-1}\left(\frac{\theta_{0}}{\sqrt{3} \sqrt{\Lambda }}\right)}{\sqrt{3} \sqrt{\Lambda } (w+1)}.
\end{equation}
For $\Lambda < 0$, this integral is finite and real if and only if $\theta_{0} = 0$. Therefore, for this choice of initial condition, this universe model will exhibit finite-time, type 0 singularities. Further, the integral diverges if $\theta_{0} = \pm \sqrt{3} \sqrt{\Lambda}$, which, as mentioned above the positive root is that of the de Sitter expansion. Therefore, a finite-time, type 0 singularity can be avoided by choosing the initial condition to be that of the de Sitter expansion.

\section{FLRW Models With Bulk Viscosity}
We now briefly consider $k=0$ FLRW models with bulk viscosity. Bulk viscous models are typically used in early-universe cosmological models. The role of the bulk viscosity is to simply add a pressure term in the perfect-fluid energy-momentum tensor. The perfect-fluid energy momentum tensor which is imposed by the symmetry of FLRW models for fluid density $\mu$, pressure $p$, and four-velocity vector $u_{a}$ is given by
\begin{equation}
T_{ab} = \left(\mu + p \right) u_{a} u_{b} + g_{ab} p.
\end{equation}
By adding a bulk viscosity term, this energy-momentum tensor is now \cite{isk1}
\begin{equation}
T_{ab} = \left(\mu + p\right) u_{a} u_{b} + g_{ab} p - \xi \theta h_{ab},
\end{equation}
where $h_{ab}$ is the projection tensor, and $\xi$ is the bulk viscosity, and $p = w \mu$. 

Following \cite{0264-9381-12-8-015}, we assume an equation of state
\begin{equation}
\xi = \xi_{0} \mu^{m},
\end{equation}
where $m$ is a constant and $\xi_{0}$ is a positive constant. 

With this definition in mind, let us define the total effective pressure as
\begin{equation}
p = \tilde{p} - \xi \theta,
\end{equation}
where $\tilde{p}$ is defined as $\tilde{p} = w \mu$.
Then, by simply substituting this definition of $p$ into Eq. \eqref{eq:friedmann1}, we see that Eq. \eqref{eq:raych1} becomes
\begin{equation}
\label{eq:newraych1}
\dot{\theta} = \frac{1}{2} \left[\theta  \kappa  3^{1-m} \xi_{0} \left(\frac{\theta ^2-3 \Lambda }{\kappa }\right)^m-\theta ^2 (w+1)+3 \Lambda  (w+1)\right].
\end{equation}

It is not possible to apply Osgood's criterion in a general sense to Eq. \eqref{eq:newraych1}, that is, for an arbitrary $m$ in Eq. \eqref{eq:newraych1}. We will therefore consider only specific values for $m$ in our analysis. In fact, the choice of $m$ is very important and can affect the overall dynamics of general FLRW models. This point was extensively studied in \cite{1986PhLB..180..335B, 1987PhLB..183..285B, 1988NuPhB.310..743B, 1990PhLB..235...40B, 0264-9381-12-8-015, belkhat}. We will only consider the cases $m=0$ and $m=1$ in what follows as these are physically relevant according to \cite{0264-9381-12-8-015} and \cite{belkhat}.

\subsection{Case: $m=0$}
This case corresponds to the case of a constant bulk viscosity. The Raychaudhuri equation \eqref{eq:newraych1} becomes
\begin{equation}
\label{eq:raych0}
\dot{\theta} = -\frac{\theta ^2}{2}+\frac{3 \theta  \kappa  \xi_{0}}{2}+\frac{3 \Lambda }{2}-\frac{\theta ^2 w}{2}+\frac{3 \Lambda  w}{2},
\end{equation}
where we will use $\theta(t_{0}) = \theta_{0}$ as an initial condition.

Then, Osgood's criterion becomes
\begin{equation}
t_s - t_0 = \int_{\theta_{0}}^{\infty}  -\frac{2 ds}{\left(1+w\right)s^2 - 3 \left(1+w\right) \Lambda - 3 s\kappa \xi}.
\end{equation}
Evaluating this integral, we find that it is equal to
\begin{equation}
\label{eq:integral2}
t_s - t_0 = \frac{2 \log \left|\frac{\sqrt{3} \sqrt{3 k^2 \xi_0^2+4 \Lambda  (w+1)^2}}{3 k \xi_0-2 \theta_0(w+1)}+1\right|-\log \left|\frac{\sqrt{3} \sqrt{3 k^2 \xi_0^2+4 \Lambda  (w+1)^2}}{2 \theta_0 (w+1)-3 k \xi_0}+1\right|}{\sqrt{3} \sqrt{3 k^2 \xi_0^2+4 \Lambda  (w+1)^2}}.
\end{equation}

This integral only diverges for
\begin{equation}
\label{eq:initcond1}
\theta_{0} = \frac{3 \kappa \xi_0 \pm \sqrt{3} \sqrt{4(1+w)^2 \Lambda + 3 \kappa^2 \xi_0^2}}{2(1+w)}.
\end{equation}
Therefore, only for the initial condition as given in Eq. \eqref{eq:initcond1}, does a unique solution exist to Eq. \eqref{eq:raych0} where there is no finite-time singularity where $\theta \to \infty$.
Finite-time singularities therefore exist, for initial conditions $\theta_0$ (assuming $-1 < w \leq 1$, $\xi_0 > 0$, $\Lambda >0$) if
\begin{equation}
\theta_0>\frac{1}{2} \sqrt{\frac{9 k^2 \xi_0^2+12 \Lambda +12 \Lambda  w^2+24 \Lambda  w}{(w+1)^2}}+\frac{3 k \xi_0}{2 (w+1)}.
\end{equation}

\subsection{Case: $m=1$}
The Raychaudhuri equation \eqref{eq:newraych1} becomes
\begin{equation}
\label{eq:raychbulk}
\dot{\theta} = \frac{1}{2}\left(\theta^2-3\Lambda\right)\left(-1-w+\theta \xi_{0}\right), \quad -1 < w \leq 1, \quad \Lambda > 0.
\end{equation}
%
Specifically, we now wish to apply Osgood's criterion to the initial value problem
\begin{eqnarray}
\dot{\theta} &=&  \frac{1}{2}\left(\theta^2-3\Lambda\right)\left(-1-w+\theta \xi_{0}\right), \\
\theta(t_0) &=& \theta_{0}.
\end{eqnarray}

Osgood's criterion then requires us to check the convergence/divergence behaviour of
\begin{equation}
\label{eq:integral2}
t_s - t_0 = \int_{\theta_{0}}^{\infty}  \frac{2 ds}{\left(s^2-3 \Lambda \right) (s \xi_{0}-w-1)}.
\end{equation}
Evaluating this integral, we find that it is equal to
\begin{eqnarray}
t_s  - t_0 = -\frac{2 \sqrt{3} (w+1) \coth ^{-1}\left(\frac{\theta_0}{\sqrt{3} \sqrt{\Lambda }}\right)-3 \sqrt{\Lambda } \xi_0 \left(2 \log |-\xi_0|+\log \left|\theta_0^2-3 \Lambda \right|-2 \log |-\xi_0 \theta_0+w+1|\right)}{3 \sqrt{\Lambda } \left(-3 \Lambda  \xi_0^2+w^2+2 w+1\right)}.
\end{eqnarray}

The integral in \eqref{eq:integral2} has finite real values if 
\begin{equation}
\left\{0 < \Lambda < \frac{1+2w+w^2}{3\xi_{0}^2}\right\} \cap \left\{\theta_{0} > \sqrt{3} \sqrt{\Lambda}\right\},
\end{equation}
or
\begin{equation}
\left\{\Lambda \geq \frac{1 + 2w + w^2}{3 \xi_{0}^2} \right\} \cap \left\{\theta_{0} > \sqrt{3} \sqrt{\Lambda}\right\}.
\end{equation}

Now, the integral in \eqref{eq:integral2} diverges if
\begin{equation}
\theta_{0} \in \left\{\sqrt{3} \sqrt{\Lambda}\right\} \cup \left\{  \pm \sqrt{1 + 3 \Lambda} \right\} \cup \left\{\frac{1+w}{\xi_{0}} \right\}.
\end{equation}
Therefore, as this calculation shows, the universe model avoids a singularity where $\theta \to \infty$ in finite time under the choice of any of these initial conditions, since with these choices the integral in Eq. \eqref{eq:integral2} diverges.

For the sake of completeness, we will also evaluate the integral in \eqref{eq:integral2} for the case of a negative cosmological constant, $\Lambda < 0$. In fact, let us define $\Lambda = -\lambda^2$, where $\lambda \neq 0 \in \mathbb{R}$. We then find that the integral in Eq. \eqref{eq:integral2} has the value
\begin{dmath}
t_s - t_0 = \frac{-\frac{\sqrt{3} \pi  \lambda  (w+1)}{\left| \lambda \right| }+3 \lambda  \xi_0 \left(2 \log|-\xi_0|-2 \log |w-\xi_0 \theta_0+1|+\log \left|3 \lambda ^2+\theta_0^2\right|\right)+2 \sqrt{3} (w+1) \tan ^{-1}\left(\frac{\theta_0}{\sqrt{3} \lambda }\right)}{3 \lambda  \left(3 \lambda ^2 \xi_0^2+w^2+2 w+1\right)}.
\end{dmath}
This has finite real values if
\begin{equation}
\theta_0 > \frac{1+w}{\xi_0}.
\end{equation}
Therefore, for these values of $\theta_0$, the universe model will admit finite-time singularities.

Further, the integral in Eq. \eqref{eq:integral2} will diverge if $\theta_0$ is chosen such that
\begin{equation}
\theta_{0} = \frac{1+w}{\xi_0}.
\end{equation}
For this choice of $\theta_0$, the universe model will not admit any finite-time singularities.


\section{FLRW Models With a Chaplygin Gas}
In this section, we attempt to use Osgood's criterion to determine the existence of finite-time, big crunch singularities in FLRW models with what is known as a Chaplygin gas. Unlike barotropic matter, the Chaplygin gas satisfies an exotic equation of state  \cite{elliscosmo},
\begin{equation}
\label{eq:chap1}
p = -\frac{A}{\mu^{\alpha}}, \quad A \in \mathbb{R}, \quad  0 < \alpha \leq 1,
\end{equation}
where $A < 0$ describes an anti-Chaplygin gas, and $A > 0$ describes a Chaplygin gas. Finite-time singularities as described above have been studied a number of times in the literature in the context of Chaplygin gas universes \cite{2016JCAP...01..040B,2015arXiv151202664C, 2015EPJC...75..244M, 2015AIPC.1647...10C, 2015OAP....28..137P, 2014AIPC.1606...79K, 2014JGP....80...58K, 2014PhRvD..89f4016B, 2013IJTP..tmp..474S, 2013Ap&SS.347..433R, 2013IJMPD..2250067M, 2011JETP..112..784B, 2010GReGr..42..745P, 2009PhRvD..79l4035B,2009PhRvD..79f3521D,2009MPLA...24..541Z, 2008PhRvD..78f4064G, 2008GrCo...14..205Y, 2008PhRvD..77j7303P, 2008PhLB..659....6P, 2008PhLB..659....1B, 2008IJMPD..17.2269B, 2007PhLB..654...51B, 2006CQGra..23.3195C, 2005PhRvD..72j3518G, 2005PhRvD..72f3511S, 2005JCAP...05..005B}. 

Using Eq. \eqref{eq:chap1} in Eqs. \eqref{eq:raych1} and \eqref{eq:friedmann1}, we obtain the following version of Raychaudhuri's equation:
\begin{equation}
\label{eq:chap2}
\dot{\theta} = \frac{1}{2} \left[3 \left(3^{\alpha } A \kappa \left(\frac{\theta ^2-3 \Lambda }{\kappa}\right)^{-\alpha }+\Lambda \right)-\theta ^2\right].
\end{equation}

For simplicity, we will consider the case $\alpha = 1$, thereby, obtaining
\begin{eqnarray}
\label{eq:chap3}
\dot{\theta} &=& -\frac{\left(\theta ^2-3 \Lambda \right)^2-9 A \kappa^2}{2 \left(\theta ^2-3 \Lambda \right)}, \\
\theta(t_0) &=& \theta_{0}.
\end{eqnarray}


Therefore, applying Osgood's criterion, we are to test the divergence of
\begin{equation}
\label{eq:chaptemp}
\int_{\theta_{0}}^{\infty} \frac{2 \left(s ^2-3 \Lambda \right)}{9 A \kappa^2-\left(s ^2-3 \Lambda \right)^2} ds.
\end{equation}
Evaluating this integral, we get
\begin{equation}
\label{eq:chap4}
t_s - t_0 = -\frac{1}{2} \pi  \sqrt{\frac{1}{3 \sqrt{A} \kappa-3 \Lambda }}+\frac{\tan ^{-1}\left(\frac{\theta_0}{\sqrt{3} \sqrt{\sqrt{A} \kappa-\Lambda }}\right)}{\sqrt{3} \sqrt{\sqrt{A} \kappa-\Lambda }}-\frac{\tanh ^{-1}\left(\frac{\theta_0}{\sqrt{3} \sqrt{\sqrt{A} \kappa+\Lambda }}\right)}{\sqrt{3} \sqrt{\sqrt{A} \kappa+\Lambda }}.
\end{equation}

Real and finite solutions corresponding to Eq. \eqref{eq:chap4} for the case $A > 0$, that is for the Chaplygin gas, occur for $\Lambda > 0$ and
\begin{equation}
\label{eq:chap6}
\sqrt{A} \kappa > \Lambda, \quad 0 < \theta_{0} < \sqrt{3} \sqrt{\sqrt{A} \kappa + \Lambda}.
\end{equation}

Finally, divergent solutions corresponding to Eq. \eqref{eq:chap4} occur when
\begin{equation}
\label{eq:chap7}
\theta_{0} = \sqrt{3} \sqrt{ \sqrt{A} \kappa + \Lambda}.
\end{equation}

As can be seen from the above calculations, finite-time singularities where $\theta \to \infty$ occur for the conditions as described in Eq. \eqref{eq:chap6}, while such a singularity is avoided if $\theta_{0}$ is chosen such that Eq. \eqref{eq:chap7} is satisfied for $A > 0$. Therefore, the finite-time singularity is avoided for a Chaplygin gas with such an initial condition. 

Considering the case of an anti-Chaplygin gas, where $A < 0$, we will apply Osgood's criterion by setting $A = -a^2$, where $a \neq 0 \in \mathbb{R}$. One can see looking at the expression in Eq. \eqref{eq:chap4}, no real solutions exist irrespective of the choice of principal function. Therefore, we are unable to determine singularity behaviour for an anti-Chaplygin gas. The physical interpretation of this result is perhaps that the expansion never diverges in this case.

A further interesting case occurs in the case of a Chaplygin gas, where $A > 0$, but with a vanishing cosmological constant, $\Lambda = 0$. In this case, applying Osgood's criterion to Eq. \eqref{eq:chaptemp}, we obtain:
\begin{equation}
\label{eq:chap8}
t_s - t_0 = \frac{\tan ^{-1}\left(\frac{\theta_0}{\sqrt{3} \sqrt[4]{A} \sqrt{\kappa}}\right)}{\sqrt{3} \sqrt[4]{A} \sqrt{\kappa}}-\frac{\tanh ^{-1}\left(\frac{\theta_0}{\sqrt{3} \sqrt[4]{A} \sqrt{\kappa}}\right)}{\sqrt{3} \sqrt[4]{A} \sqrt{\kappa}}.
\end{equation}
As can be seen Eq. \eqref{eq:chap8} has finite and real values if
\begin{equation}
0 < \theta_0 < \sqrt{3} \sqrt{ \sqrt{A} \kappa}.
\end{equation}
Therefore, for these values of $\theta_0$, there exist finite-time singularities. Finite-time singularities for a Chaplygin gas with vanishing cosmological constant were demonstrated in \cite{Kamenshchik2001265}.

Further, finite-time singularities can be avoided if $\theta_0$ is chosen such that
\begin{equation}
\theta_0 \in \left\{0\right\} \cup \left\{\sqrt{3} \sqrt{ \sqrt{A} \kappa}\right\}.
\end{equation}

\section{On Sudden Singularities}
We now given an example of a situation where Osgood's criterion can be used to prove the existence of Type II / Sudden Singularities as described in Table \ref{table:table1} above. We will consider an anti-Chaplygin gas, with $\alpha = 1$ as per Eq. \eqref{eq:chap1}. We will further assume that the cosmological constant vanishes, $\Lambda = 0$. In the case of a sudden singularity, the pressure diverges, that is, $p \to \infty$, while the energy density and expansion vanish. To apply Osgood's criterion to this situation, we use Eqs. \eqref{eq:endensity1}, \eqref{eq:friedmann1}, and \eqref{eq:chap1} to formulate a new initial value problem for the pressure, $p(t)$:
\begin{eqnarray}
\label{eq:sudden1}
\dot{p} &=& -\left(\frac{a^2 \kappa}{p}\right)^{-1/2} \sqrt{3} \kappa \left(a^2 + p^2\right), \\
\label{eq:sudden2}
p(t_0) &=& p_{0},
\end{eqnarray}
where for simplicity, we have denoted the anti-Chaplygin gas constant $A$, by $A = -a^2$, where $a \neq 0 \in \mathbb{R}$.  

Applying Osgood's criterion to the initial value problem \eqref{eq:sudden1}-\eqref{eq:sudden2}, we are to evaluate
\begin{equation}
t_s - t_0 = \int_{\theta_0}^{\infty} - \frac{\left( \frac{a^2 \kappa}{s}\right)^{1/2}}{\sqrt{3} \left(a^2 \kappa + \kappa s^2\right)} ds.
\end{equation}
We find that
\begin{dmath}
\label{eq:sudden2}
t_s - t_0 = -\frac{\sqrt{a} \left[-\log \left|\sqrt{2} \sqrt{a} \sqrt{p_0}+a+p_0\right|+\log \left|-\sqrt{2} \sqrt{a} \sqrt{p_0}+a+p_0\right|-2 \tan ^{-1}\left(\frac{\sqrt{2} \sqrt{p_0}}{\sqrt{a}}+1\right)+2 \tan ^{-1}\left(1-\frac{\sqrt{2} \sqrt{p_0}}{\sqrt{a}}\right)\right]}{2 \sqrt{6} a\sqrt{\kappa}}-\frac{\pi +\log \left|3-2 \sqrt{2}\right|}{2 \sqrt{6} \sqrt{a \kappa}}.
\end{dmath}

One can easily confirm that Eq. \eqref{eq:sudden2} is real and finite for all $p_0 \in \mathbb{R}$.  Therefore, for all $p_0 \in \mathbb{R}$, we have that the universe model will exhibit a finite-time sudden singularity. 

\section{Conclusions}
In this paper, we have applied Osgood's criterion to detect finite-time singularities in a spatially flat FLRW universe in the context of a perfect fluid, a perfect fluid with bulk viscosity, and a Chaplygin and anti-Chaplygin gas. In particular, we applied Osgood's criterion to demonstrate singularity behaviour for Type 0/big crunch singularities as well as Type II/sudden singularities. We have shown that in each case the choice of initial conditions is important as a certain number of initial conditions leads to finite-time singularities, while other precise choices of initial conditions which depend on the cosmological matter parameters and the cosmological constant can avoid a finite-time singularity. 

As we have shown, Osgood's criterion provides a powerful and yet simple way of deducing the existence of these singularities, and also interestingly enough, provides clues of how to eliminate singularities from certain cosmological models. For this reason, we believe that the work here is unique, and could be of interest in future cosmological studies.


\newpage
\bibliographystyle{ieeetr} 
\bibliography{sources}

\end{document}